\documentclass[12pt]{article}
\usepackage{latexsym,epsfig,amssymb,amsmath}
\usepackage{color}
\usepackage{cancel}
\newcommand{\be}{\begin{equation}}
\newcommand{\ee}{\end{equation}}
\newcommand{\ben}{\begin{displaymath}}
\newcommand{\een}{\end{displaymath}}
\newcommand{\bea}{\begin{eqnarray}}
\newcommand{\eea}{\end{eqnarray}}

\newcommand{\bean}{\begin{eqnarray*}}
\newcommand{\eean}{\end{eqnarray*}}
\newcommand{\beqs}{\begin{eqnarray}}
\newcommand{\eeqs}{\end{eqnarray}}

\newcommand{\mathon}{\mathversion{bold}}
\newcommand{\mathoff}{\mathversion{normal}}

\makeatletter
\@addtoreset{equation}{section}
\makeatother

\begin{document}

\thispagestyle{empty}

\begin{flushright}\small
\end{flushright}


\bigskip
\bigskip

\mathon
\vskip 10mm

\begin{center}

  {\Large{\bf Rotating D0-branes
  and consistent truncations\\[1ex] of supergravity}}

\end{center}

\mathoff


\vskip 6mm

\begin{center}
{\bf Andr\'es Anabal\'on$^{a,b}$, Thomas Ortiz$^b$, and Henning Samtleben$^b$}\\[3ex]

$^a$\,{\em Departamento de Ciencias, Facultad de Artes Liberales y\\
Facultad de Ingenier\'{\i}a y Ciencias, Universidad Adolfo Ib\'{a}\~{n}ez,\\ Av. Padre Hurtado 750, Vi\~{n}a del Mar, Chile.} \\
\vskip 4mm

$^b$\, {\em Universit\'e de Lyon, Laboratoire de Physique, UMR 5672, CNRS\\
\'Ecole Normale Sup\'erieure de Lyon\\
46, all\'ee d'Italie, F-69364 Lyon cedex 07, France}\\

\end{center}

\vskip1.8cm

\begin{center} {\bf Abstract } \end{center}

\begin{quotation}

\noindent
The fluctuations around the D0-brane near-horizon geometry are described by
two-dimensional $SO(9)$ gauged maximal supergravity.
We work out the $U(1)^{4}$ truncation of this theory whose
scalar sector consists of five dilaton and four axion fields. We construct the full non-linear
Kaluza-Klein ansatz for the embedding of the
dilaton sector into type IIA supergravity. This yields a consistent truncation
around a geometry which is the warped product of a two-dimensional domain wall and the sphere $S^8$.
As an application, we consider the solutions corresponding to rotating D0-branes
which in the near-horizon limit approach AdS$_2 \times {\cal M}_8$ geometries,
and discuss their thermodynamical properties.
More generally, we study the appearance of such solutions in the presence of
non-vanishing axion fields.

 \end{quotation}

\newpage
\setcounter{page}{1}


\newpage


\section{Introduction}

The standard AdS/CFT correspondence~\cite{Maldacena:1997re},
is based on maximally supersymmetric string/brane configurations
whose near horizon geometry factorizes into the direct product of and
Anti de Sitter (AdS) space and a sphere.
Its generalization to non-conformal D$p$-branes~\cite{Itzhaki:1998dd}
proposes a holographic description for the
maximally supersymmetric $(p+1)$-dimensional
Yang-Mills theory (non-conformal for $p\not=3$).
In this latter case, the relevant near-horizon geometry is the warped
product of a $(p+2)$-dimensional domain wall and the sphere $S^{8-p}$.
Relatively few tests of these non-conformal dualities have been carried out,
and only more recently the techniques of holographic renormalization
have been developed systematically also in
the non-conformal context~\cite{Wiseman:2008qa,Kanitscheider:2008kd}.

In the supergravity approximation, these so-called domain wall/quantum field theory (DW/QFT)
correspondences~\cite{Boonstra:1998mp,Behrndt:1999mk,Bergshoeff:2004nq}
imply the duality between certain gauged supergravities supporting
half-maximal domain-wall solutions, and certain subsectors of the non-conformal quantum field theories.
A key issue for the applicability of the lower-dimensional effective actions in
this context is the question of their consistent embedding into the full
ten-dimensional theory. Consistent truncations of higher-dimensional supergravities
are highly constrained and relatively rare.
For the maximal (AdS) cases, consistency of the truncations on
AdS$_4\times S^7$ and AdS$_7\times S^4$  have been shown in
\cite{deWit:1986iy,Nicolai:2011cy} and~\cite{Nastase:1999kf}, respectively,
whereas the consistent truncation on AdS$_5\times S^5$ so far has only been shown
for certain subsectors~\cite{Cvetic:1999xp,Cvetic:1999xx,Cvetic:2000nc}.
For the non-conformal cases, in which the geometry is the warped
product of a domain wall and a sphere, the results about consistent truncations
are more sporadic. They include most notably~\cite{Cvetic:1999pu} in which consistent truncations
are obtained for domain-wall supergravities obtainable from particular limits of AdS supergravities,
and~\cite{Cvetic:2000zu} for reductions on small spheres.

The subject of this letter are compactifications of the IIA theory on the eight-sphere $S^8$,
describing fluctuations around the D0 brane near-horizon geometry. The low-energy effective
theories are truncations of the two-dimensional maximally supersymmetric $SO(9)$ gauged supergravity,
recently constructed in~\cite{Ortiz:2012ib}. The dual field theory in this case
is the supersymmetric matrix quantum mechanics~\cite{deWit:1988ig}
which itself has been proposed as a non-perturbative definition of
M-theory~\cite{Banks:1996vh}.
More precisely, we study the (non-supersymmetric) truncation of $SO(9)$ supergravity to singlets
under its Cartan subgroup $U(1)^4$. Its scalar sector consists of five dilaton and four axion fields.
In analogy to the well-known results for the AdS supergravities~\cite{Cvetic:1999xp}, we
construct the non-linear Kaluza-Klein ansatz for the embedding of the
full dilaton sector into type IIA supergravity.
Contrary to the AdS case, this ansatz requires a non-constant ten-dimensional dilaton field which
gives rise to an additional dilaton in two dimensions with non-vanishing contribution
to the domain wall ground state.
Moreover, this $p=0$ case is special in that there is no Einstein frame in two dimensions such
that many of the generic reduction ansaetze~\cite{Cvetic:1999xp,Cvetic:2000zu} in fact degenerate.

As an application, we construct the solutions corresponding to
rotating D0-branes which in the limit of large brane charges
approach AdS$_2 \times {\cal M}_8$ geometries. More generally, we
study the appearance of such solutions (in particular in the
presence of non-vanishing axion fields). Their existence is due to
non-vanishing flux of the two-dimensional vector fields.

The appearance of an AdS$_2$ in the near extremal limit allows to
use Sen's entropy function formalism \cite{Sen:2005wa} to compute the
two-dimensional entropy of this configuration. We find that the entropy is proportional 
to the value of the two-dimensional dilaton 
(which precludes the existence of the Einstein
frame as otherwise the Einstein-Hilbert Lagrangian would be a total derivative). 
The result exactly
matches the ten-dimensional result given by the Bekenstein-Hawking
area law. Remarkably, the fact that two-dimensional black
holes have zero-dimensional horizons is not an obstruction for Sen
formalism to work properly.

The outline of the paper is as follows: in the second section the
$SO(9)$ theory is briefly reviewed and the $U(1)^4$ truncation
discussed. The third section displays the explicit embedding of the
truncation in type IIA supergravity. The fourth section recalls the
rotating D$0$-brane solution of \cite{Cvetic:1999xp} as well as its
thermodynamical properties from the ten-dimensional point of view.
In the fifth section the two-dimensional entropy function formalism
is applied to the limit of large brane charges of the rotating
D$0$-brane. The attractor mechanism is shown to work and the 
ten-dimensional result of the previous section is exactly recovered. In
the last section, the axions are included in the truncation and its
ten-dimensional embedding is sketched.


\section{${U(1)}^{4}$ truncation of $SO(9)$ supergravity}
\label{sec:truncation}

$SO(9)$ supergravity is a maximally supersymmetric theory in two dimensions.
It has been constructed in~\cite{Ortiz:2012ib} by gauging the $SO(9)$ subgroup of
the global $SL(9)$ symmetry that is manifest after reduction of eleven-dimensional
supergravity on a nine-torus $T^9$.
A consistent truncating of the theory is given by truncating the theory to singlets
under the $U(1)^4$ Cartan subgroup of $SO(9)$. This is in complete analogy to
the truncations of higher-dimensional maximal supergravities considered in~\cite{Cvetic:1999xp}:
the $U(1)^4$ truncation of $D=4$, $SO(8)$ supergravity and the $U(1)^3$ and $U(1)^2$
truncations of $D=5$, $SO(6)$ and $D=7$, $SO(5)$ supergravity, respectively.
For $D=2$, $SO(9)$ supergravity the truncation to $U(1)^4$ singlets reduces the coordinates
of the scalar target space to five dilaton and four axion fields. Moreover, the bosonic Lagrangian
carries four abelian gauge fields and four auxiliary scalar fields.

\paragraph{$SO(9)$ Action}

We start from the action given in  \cite{Ortiz:2012ib}, formula $(4.2)$.
Its bosonic part is a dilaton-gravity coupled non-linear sigma model with
128-dimensional target
space $\left(SL(9) \ltimes T_{84} \right)/ SO(9)$ and Wess-Zumino term. Thus $80$ scalar fields are
parametrized by group-valued $SL(9)$ matrices ${{\cal{V}}_{m}}^{\alpha}$ with coset redundancy
\bea
{{\cal{V}}_{m}}^{\alpha} &\rightarrow& {{\cal{V}}_{m}}^{\beta}\,\Lambda_\beta{}^\alpha\;,\qquad
\Lambda\in SO(9)\;.
\eea
The remaining $84$ scalars fields are labeled as $\phi^{klm}=\phi^{[klm]}$ with the indices running
from $1$ to $9$. The $SO(9)$ symmetry acting by left multiplication on ${{\cal{V}}_{m}}^{\alpha}$
and rotating the scalars $\phi^{klm}$ in the ${\bf 84}$ representation is gauged by
$36$ vector fields $A_{\mu}^{kl}=A_{\mu}^{[kl]}$. In turn, these couple to $36$ auxiliary fields
$Y^{kl}=Y^{[kl]}$ in the Lagrangian. Finally let us precise that our signature is $(+-)$.
The explicit form of the Lagrangian is
\begin{equation}
\begin{aligned}
 {\cal{L}}=&  -\frac{1}{4} e \rho \, R + \frac{1}{4} e \rho \,  {\cal{P}}^{\alpha \beta}_{\mu} {\cal{P}}^{\mu\,\alpha \beta}
+ \frac{1}{12} \, e \rho^{1/3} \, {{\cal{V}}_{klm}}^{[\alpha \beta \gamma]} {{\cal{V}}_{npq}}^{[\alpha \beta \gamma]}
D_{\mu} \phi^{klm} D_{\mu} \phi^{npq} \\
& + \frac{1}{648} \varepsilon^{\mu \nu} \varepsilon_{klmnpqrst} \, \phi^{klm} D_{\mu} \phi^{npq} D_{\nu} \phi^{rst}
- \frac{g}{4} \varepsilon^{\mu \nu} {{F}}_{\mu \nu}^{kl}  \, Y^{kl}
- e \, V_{\text{pot}}({\cal V}, \phi, Y)\;,\\
\end{aligned}
\label{LSO9}
\end{equation}
with the scalar currents defined by
\bea
J_{\mu}^{\alpha \beta} &\equiv&
{\cal{V}}^{-1 \alpha k} \left( \partial_{\mu} {{\cal{V}}_{k}}^{\beta} + gA_{\mu}^{kl} {{\cal{V}}_{l}}^{\beta}  \right)
~\equiv~ {\cal Q}_{\mu}^{[\alpha \beta]}+{\cal P}_{\mu}^{(\alpha \beta)}\;,
\nonumber\\
D_{\mu} \phi^{klm} &=& \partial_{\mu} \phi^{klm} - 3 \, g\,A_{\mu}^{p[k} \phi^{lm]p}
\;.
\eea
The first term of the Lagrangian carries the two-dimensional Ricci scalar $R$,
the determinant of the zweibein
$e=\sqrt{|{\rm det}\,g_{\mu\nu}|}$, and the dilaton field $\rho$\,.
$F_{\mu\nu}^{kl}$ denotes the standard non-abelian $SO(9)$ Yang-Mills
field strength of the vector fields.
Also, we have introduced an explicit coupling constant $g$\,.
We refrain from spelling out the somewhat lengthy explicit expression
for the scalar potential $V_{\text{pot}}$ which has been given in~\cite{Ortiz:2012ib}.
Let us just mention, that it can be expanded as an eighth order polynomial in
the scalars $\phi^{klm}$ with the lowest order given by
\bea
 V_{\rm pot}
 &=&   \frac{g^2}{8} \, \rho^{5/9} \left( 2 \, {\rm tr}[{{M}}{{M}}]  - \left({\rm tr}\,{{M}}\right)^2
 \right)
 + g^2 \rho^{-13/9} {{M}}^{km} {{M}}^{ln}
 \,Y_{kl}  Y_{mn}
~+~{\cal O}(\phi^2)
\;,
\label{pot_exp}
\eea
with the matrix ${M}$ defined by ${M}\equiv ({\cal V}{\cal V}^T)^{-1}$\,.
Modulo the dilaton prefactor, the first term of the potential is precisely the contribution
expected from a sphere compactification, cf.~\cite{Cvetic:1999xx,Cvetic:2000dm}.

\paragraph{${U(1)}^{4}$ truncation}

We will now consider the consistent truncation of the Lagrangian (\ref{LSO9}) to
singlets under the $U(1)^{4}$ Cartan subgroup of the gauge group $SO(9)$.
Explicitly, we choose a parametrization in which gauge fields and auxiliary scalars reduce to
\bea
{A_{\mu}}^{kl} \equiv  A_{\mu}^{a} \, {T_{a}}^{kl}
\;,\qquad
Y^{kl} \equiv \frac{\rho}{4}\,y^{a} \,{T_{a}}^{kl}\;,\qquad
a=1, \dots, 4
\;,
\eea
with generators
\bea
T_1{}^{kl} \equiv 2 \delta_1^{[k}\delta_2^{l]}
\;,\quad
T_2{}^{kl} \equiv 2 \delta_3^{[k}\delta_4^{l]}
\;,\quad
T_3{}^{kl} \equiv 2 \delta_5^{[k}\delta_6^{l]}
\;,\quad
T_4{}^{kl} \equiv 2 \delta_7^{[k}\delta_8^{l]}
\;.\eea
Correspondingly, the scalar matrix ${\cal V}$ reduces to
\begin{eqnarray}
{\cal{V}}&=& \text{exp}\left( v_{a} h^{a} \right) \;,
 \label{truncV}\\[.5ex]
&&
h^{1}\equiv\text{diag}\left(1,1,0,0,0,0,0,0,-2\right), \; h^{2}\equiv\text{diag}\left(0,0,1,1,0,0,0,0,-2\right) , \nonumber\\
&& h^{3}\equiv\text{diag}\left(0,0,0,0,1,1,0,0,-2\right), \; h^{4}\equiv\text{diag}\left(0,0,0,0,0,0,1,1,-2\right),
\nonumber
\end{eqnarray}
parametrized by four (dilaton) scalar fields $v_\alpha$\,. From the 84 (axion) scalars $\phi^{klm}$,
four survive the truncation:
\bea
\phi^1\equiv\phi^{129}\;,\quad
\phi^2\equiv\phi^{349}\;,\quad
\phi^3\equiv\phi^{569}\;,\quad
\phi^4\equiv\phi^{789}\;,
\label{truncphi}
\eea
with all other components vanishing.
The resulting action is obtained by plugging this truncation into (\ref{LSO9}) and takes the form
\bea
 {\cal{L}}&= &  -\frac{1}{4} e \rho \, R + \frac{1}{2} e \rho \sum_{a} \partial_{\mu} u_{a} \,  \partial^{\mu} u_{a}
 + \frac{1}{2} e \rho^{1/3} X_{0}^{-1} \sum_{a=1}^{4} X_{a}^{-2} \left( \partial_{\mu} \phi^{a} \right) \left( \partial^{\mu} \phi^{a} \right)
\nonumber\\
&&{}- \frac{\rho}{8}\,g \, \varepsilon^{\mu \nu} F_{\mu\nu}^{a} \; y^{a}   - e \,V_{\text{pot}}\;,
\label{Ltrunc}
\eea
where we have defined
\bea
X_{a} \equiv e^{-2 v_{a}} \equiv e^{-2\,A_{ab} u_{b}}\;,\qquad
X_{0} \equiv (X_1X_2X_3X_4)^{-2} \;,
\eea
with the matrix
\bea
A = {\footnotesize \begin{pmatrix}
  1/6 & -1/\sqrt{2} & -1/\sqrt{6} & -1/(2\sqrt{3})\\
  1/6 & 0 & 0 & \sqrt{3}/2\\
  1/6 & 0 & \sqrt{2/3} & -1/(2\sqrt{3})\\
  1/6 & 1/\sqrt{2} & -1/\sqrt{6} &  -1/(2\sqrt{3})\\
 \end{pmatrix}}
 \;,
 \nonumber
\eea
and the abelian field strengths
$F_{\mu\nu}^{a} \equiv 2 \, \partial_{[\mu} A_{\nu]}^{a}$\,.

The scalar potential $V_{\text{pot}}$ in (\ref{Ltrunc})
can be obtained upon evaluating the explicit expressions given in~\cite{Ortiz:2012ib}
for the truncation (\ref{truncV}), (\ref{truncphi}).\footnote{
We use the occasion to correct a crucial typo in~\cite{Ortiz:2012ib} in the definition of the scalar tensors
which are the building blocks of the scalar potential.
In equation (4.18), the second term on the r.h.s.\ in the definition of the tensor $b^a$ should
carry a factor of $-\frac1{288}$ instead of $\frac1{144}$\,.}
After some computation, this leads to the expression
\bea
V_{\text{pot}} &=& g^{2} \, \rho^{5/9} \,\Big[ \frac{1}{8} \,\Big({X_{0}}^{2} - 8 \sum_{a<b} X_{a} X_{b} - 4 X_{0} \sum_{a} X_{a} \Big)
+ \frac{1}{2} \, \rho^{-2/3}  \sum_{a} X_{a}^{-2}  \left( X_{0} - 4 X_{a} \right) (\phi^{a})^{2} \nonumber\\
&&\qquad\quad{}  + 2 \, \rho^{-4/3} \sum_{a<b}X_{a}^{-2} X_{b}^{-2} (\phi^{a})^{2} (\phi^{b})^{2}
+ \frac18 \, \rho^{-2} \sum_{a} X_{a} \,  \Big(\rho \, y^{a} + 8 \,  \prod_{b \neq a} \phi^{b} \Big)^{2} \nonumber\\
&&\qquad\quad{}  + \frac12 \, \rho^{-8/3} X_{0}^{-1} \Big( \sum_{a}  \, \rho \,y^{a} \phi^{a} + 8  \prod_{a} \phi^{a} \Big)^{2}  \, \Big]
\;,
\eea
as a fourth order polynomial in the scalars $\phi^a$\,.
Upon further field redefinition
\bea
X_{a} \equiv H_{a} \, X_{0} \;,\qquad 
\phi^{a} \equiv \frac{1}{2} \, \rho^{1/3} \, \eta_{a} \, X_{a}\,  X_{0}^{1/2}\;,
\label{defH}
\eea
the potential takes the more compact form
\bea
V_{\text{pot}} &=&   \frac{g^{2}}{8} \, \rho^{5/9} H_{0}^{-4/9}  \,\Big[ 1 - 8 \sum_{a<b} H_{a} H_{b} - 4 \sum_{a} H_{a}
+ \sum_{a}\left(1-4H_{a} \right) \eta_{a}^{2}
+\sum_{a<b} \eta_{a}^{2} \eta_{b}^{2} \nonumber\\
&&{}\qquad\qquad\qquad +\sum_{a} \eta_{a}^{-2} ( y^a  H_{a} \eta_a
+   \eta_0  )^2
+ \Big( \eta_0 + \sum_{a}  y^a H_a \eta_a   \Big)^2 \Big]
\;,
\label{Vcomp}
\eea
with $H_{0} \equiv H_1 H_2 H_3 H_4$, $\eta_0 \equiv \eta_1\eta_2\eta_3\eta_4$\,.

\paragraph{Integrating out the auxiliary fields}

The auxiliary scalar fields $y^a$ can be eliminated from the Lagrangian by virtue of their field equation
\bea
 y^{a} &=&
 -\sum_{b} {{\cal{O}}^{-1}}_{ab} \Big(  \frac{1}{2} (g e)^{-1} \, \rho^{4/9} \, \varepsilon^{\mu \nu} F_{\mu \nu}^{b}
 + 8 \,{\cal{O}}_{bb} \prod_{c \neq b} \phi^{c} \Big)\;,\nonumber\\
&&\mbox{with the matrix~~}{\cal{O}}_{ab}~ \equiv~  X_{a} X_{b} \; (\delta_{ab} + \eta_{a} \eta_{b})  \equiv  X_{a} X_{b} \; m_{ab} \;.
 \eea
Then, the vector fields acquire a two-dimensional Maxwell term together with another coupling linear in the field strengths.
\bea
{\cal{L}}&= & -\frac{1}{4} e \rho \, R + \frac{1}{2} e \rho \; \sum_{a} \left( \partial_{\mu} u_{a} \right) \left( \partial^{\mu} u_{a} \right)
+ \frac{1}{2} e  \rho^{1/3} H_{0}^{2/3} \sum_{a} H_{a}^{-2} \left(\partial_{\mu} \phi^{a} \right) \left(\partial^{\mu} \phi^{a} \right)
\nonumber\\
&&{}
- \frac{e}{16} \, \rho^{13/9} \,H_{0}^{4/9} \sum_{a,b} H_{a}^{-1} H_{b}^{-1} m^{-1}{}_{ab} \, {F_{\mu \nu}}^{a} F^{\mu \nu b}
\nonumber\\
&&{} + \frac{g}{8} \rho \, \eta_0  \sum_{a,b} \varepsilon^{\mu \nu} {F_{\mu \nu}}^{a} \,H_{a}^{-1} \eta_{b}^{-1} (1+\eta_{b}^{2}) \, m^{-1}{}_{ab}
- e \widehat{V}_{\rm pot}
\;,
\label{LtruncFF}
\eea
with the modified scalar potential given by
\bea
\widehat{V}_{\rm pot} &=& \frac{g^{2}}{8} \, \rho^{5/9} H_{0}^{-4/9} \,\Big( 1 - 8 \sum_{a<b} H_{a} H_{b} - 4 \sum_{a} H_{a}
+\frac{9 \, \eta_{0}^{2}}{1+\sum_{a} \eta_{a}^{2}}  \nonumber\\
&&{} \qquad\qquad \qquad\quad + \sum_{a} \left(1-4H_{a} \right) \eta_{a}^{2}
+\sum_{a<b} \eta_{a}^{2} \eta_{b}^{2} \Big)
\;.
\eea

\paragraph{Dilaton sector}

In the following, we will
mainly study the truncation of the Lagrangian (\ref{LtruncFF}) to the dilaton fields $\{X_a, \rho\}$, i.e.\ set the axions $\phi^a \equiv 0$\,.
The form of the Lagrangian (\ref{LtruncFF}) shows that this is a consistent further truncation of the model.
This is in contrast to the analogous model in four dimensions, where the $U(1)^4$ of $SO(8)$ supergravity gives
rise to a scalar sector of 3 dilators and 3 axions. In that case, the axions are sourced by the 
field strengths as $F\wedge F$,
such that in general they may not consistently be truncated~\cite{Cvetic:1999xp}.
Under this further truncation, the Lagrangian (\ref{LtruncFF}) reduces to
\bea
{\cal{L}}&= & -\frac{1}{4} e \rho \, R + \frac{1}{2} e \rho \; \sum_{a} \partial_{\mu} u_{a} \, \partial^{\mu} u_{a}
- \frac{1}{16} e \rho^{13/9} \; \sum_{a} X_{a}^{-2} {F_{\mu \nu}}^{a} {F^{\mu \nu}}^{a}  \nonumber\\
&&{} -\frac{1}{8}\, eg^{2}
\rho^{5/9} \,\Big( \left(X_{1} X_{2} X_{3} X_{4}\right)^{-4} - 8 \sum_{a<b} X_{a} X_{b}
 - 4 \,\frac{\sum_{a} X_{a}}{\left( X_{1} X_{2} X_{3} X_{4} \right)^{2}} \Big)  \;.
\label{Ldilaton}
\eea
This form of the Lagrangian closely resembles the analogous truncations of the maximal AdS supergravities
in $D=4, 5, 7$~\cite{Cvetic:1999xp}, with the non-trivial dilaton couplings in $\rho$ exposing the fact that
this theory supports a domain wall solution.
To analyze  the dynamics of the system, it is more convenient to pass over to the Lagrangian with auxiliary fields
$y^a$, obtained by truncation of (\ref{Ltrunc})
\bea
 {\cal{L}}&= &  -\frac{1}{4} e \rho \, R + \frac{1}{2} e \rho \sum_{a} \partial_{\mu} u_{a} \,  \partial^{\mu} u_{a}
- \frac{\rho}{8}\,g \,  \sum_a \varepsilon^{\mu \nu} F_{\mu\nu}^{a} \; y^{a}
\nonumber\\
&&{}
  -\frac{e g^{2}}{8} \, \rho^{5/9} H_{0}^{-4/9}  \,\Big( 1 - 8 \sum_{a<b} H_{a} H_{b} - 4 \sum_{a} H_{a}
   +\sum_{a} ( y^a  H_{a} )^2   \Big)
\;,
\label{LdilatonF}
\eea
in terms of the variables (\ref{defH}). Now the vector field equations state that the products
$(\rho\, y^a)$ are constant, while variation
of the Lagrangian w.r.t.\ $y^a$ expresses the field strengths as functions of the scalar fields
according to
\bea
F_{\mu \nu}^a &=& g \, e \, \varepsilon_{\mu\nu}\,\rho^{-4/9} H_{0}^{-4/9} H_{a}^{2} \; y^a
\;.
\label{eqmF}
\eea
The remaining equations of motion are given by the scalar field equations
\bea
\sum_{b} \left( \rho^{-1} \nabla^{\mu} \left( \rho \, \partial_{\mu} u_{b} \right) {A^{-1}}_{b \,a}  \right)
\qquad\qquad\qquad
\qquad\qquad\qquad
\qquad\qquad\quad
&&
 \nonumber\\
  + \, g^{2} \rho^{-4/9}  H_{0}^{-4/9} \Big(1 + 2 \, H_{a} \sum_{b\neq a} H_{b}
 + H_{a}-2\sum_{b} H_{b} -\frac{1}{2} ({y^a}H_{a} )^2 \Big) &=&0
 \;,
 \label{eqm:scalars}
\eea
the traceless part of the Einstein equations
\bea
 \rho^{-1} \nabla_{\mu} \partial_{\nu} \rho + 2 \, \sum_{a}  \partial_{\mu} u_{a} \, \partial_{\nu} u_{a}
 &=& \frac12\,g_{\mu\nu} \, \Big(
 \rho^{-1} \nabla_{\rho} \partial^{\rho} \rho + 2 \, \sum_{a}  \partial_{\rho} u_{a} \, \partial^{\rho} u_{a}
 \Big)
 \;,
 \label{Einstein1}
\eea
and suitable combinations of the dilaton and the trace part of the Einstein equations
\bea
R  &=& 2 \, \sum_{a} \left( \partial^{\mu} u_{a} \right) \left( \partial_{\mu} u_{a} \right) \nonumber\\
&&{}{-\frac{5}{18}}\,g^2 \,  \rho^{-4/9} \, H_{0}^{-4/9} \Big(
 1   - 8 \sum_{a<b} H_{a} H_{b}
- 4 \sum_{a} H_{a}  {- \frac{13}{5}} \sum_{a} (y^a H_{a})^{2}   \Big)
\;,\nonumber\\[1ex]
\rho^{-1} \nabla^\mu \partial_\mu \rho &=&
\sum_{a,b} \left( \rho^{-1} \nabla^{\mu} \left( \rho \, \partial_{\mu} u_{b} \right) {A^{-1}}_{b \,a}  \right)
+ \frac92\,g^2\,\rho^{-4/9} H_{0}^{-4/9} \Big(
  1 - 2 \sum_{a} H_{a}
 \Big)
 \;.
 \label{Einstein2}
\eea

\paragraph{Particular solutions}

The simplest solutions of the dilaton sector described above are given by assuming constant
scalars $H_a$ and a domain wall ansatz
\begin{equation}
ds^2 = e^{2 \, A(r)} dt^2 - dr^2
\;,
\label{DW}
\end{equation}
for the two-dimensional metric.
In this case, equations (\ref{eqm:scalars})
determine the auxiliary scalars $y^a$ (therefore the field strengths) in terms of
the remaining scalars as
\bea
({y^a})^2&=& 2\,H_{a}^{-2}(1+H_a)-4   + 4 \,  \sum_{b} \frac{H_{b}(H_a-1)}{H_{a}^2}  \;.
\label{detY}
\eea
The remaining equations (\ref{Einstein1}), (\ref{Einstein2}) then determine $\rho$ and the two-dimensional
metric. As special cases, we identify
\begin{itemize}
\item
the case of vanishing field strengths, i.e.~$y^a=0$. Equations (\ref{detY}) then
imply that all scalar fields
are equal $H_1=H_2=H_3=H_4\equiv {\rm H}$,
(recall that $H_a>0$),
with two distinct solutions
\bea
{\rm H} = 1\;,\quad \mbox{or} \quad {\rm H} = \frac16\;.
\eea
The first choice describes the half-supersymmetric domain-wall solution,
completed as
\bea
\rho &=& (gr)^{{9/2}}\;,\qquad A(r)~=~\frac72\,{\rm ln}\,r
\;,\qquad R~=~ \frac{35}{2} \frac{1}{r^{2}}
\;,
\label{dow}
\eea
corresponding to the ten-dimensional D0-brane near-horizon geometry.
The choice ${\rm H} = \frac16$ corresponds to a non-supersymetric
domain-wall with more complicated function $A(r)$.

\item
the AdS case: imposing a constant dilaton field
$\rho$, equation~(\ref{Einstein2}) implies
\bea
\sum_{a} H_{a}=\frac12
\;,
\label{const_rho1}
\eea
and the remaining equations of motion are solved by a two-dimensional AdS metric
\begin{equation}
\label{const_rho2}
\left\{
\begin{aligned}
 & \text{ds}^{2} = {f(r)} \, \text{dt}^{2} - \frac{1}{{f(r)}}
 \, \text{dr}^{2}  \\
&\\
& {f(r)}=-C +  g^{2} \, \frac{ \left(1+ 8 \, \sum_{a<b} H_{a} H_{b}
\right)\,r^2}{2 \, \rho^{4/9}
\, H_{0}^{4/9}}\\
&\\
 & {F_{\mu \nu}}^{a} =2 g  \, \rho^{-4/9} \left( \frac{H_{a}^{2}
\,\sqrt{H_{a}^{-1}-1} }{H_{0}^{4/9}} \right) \, e \, \epsilon_{\mu \nu} \\
&\\
& r_{\rm AdS} = \frac{\sqrt{2} \, \rho^{2/9} H_{0}^{2/9}}{g} \, \Big(1+ 8 \, \sum_{a<b} H_{a} H_{b}\Big)^{-1/2}\\
\end{aligned}
\right.
\end{equation}

Where $C$ is an integration constant. We obtain a three-parameter
family of pure AdS$_2$ solutions. The Killing spinor equations
corresponding to the Lagrangian (\ref{LSO9}) quickly show that these
solutions break all supersymmetries. While this metric is locally
AdS it clearly resembles the ($r$-$t$) section of non-rotating BTZ black
hole \cite{Banados:1992wn, Banados:1992gq} with $C$ being the mass
of the spacetime.

\end{itemize}

\section{Embedding into IIA Supergravity}

We can now describe one of our main results:
the reduction ansatz for the embedding of the two-dimensional model (\ref{Ldilaton})
into type IIA supergravity. The relevant part of the ten-dimensional Lagrangian is given by
\bea {4\pi G \cal L} &=& -\frac14 e\,R + \frac{1}{2} \, e
\,\partial_M\phi \,\partial^M\phi - \frac{1}{16} e \,e^{3 \phi}
\,F_{MN}\,F^{MN} \;. \label{L2A} \eea We split the ten-dimensional
coordinates into $\{x^M\} \rightarrow \{x^\mu, \mu_a, \sigma_a \}$,
with $\mu=0, 1$, $a=1, \dots, 4,$ in accordance with the conventions
of the previous section.
The non-linear Kaluza-Klein ansatz for the proper embedding of the two-dimensional fields
is given by generalizing the AdS reduction ansaetze from \cite{Cvetic:1999xp} to non-constant dilaton
and two-dimensional external space. The ten-dimensional
metric then is given by
\bea
ds_{10}^2 &=&  \rho^{-7/36}
\Delta^{7/8} \,ds_2^2 \nonumber\\
&&{}
- g^{-2} \, \rho^{1/4}\,\Delta^{-1/8}\,\Big(
X_0^{-1} d\mu_0^2 + \sum_{a}
X_a^{-1}\left(d\mu_a^2 + \mu_a^2\left(d\sigma_a +  g \, A^a\right)^2\Big)
\right) \;,
\label{ansatz:metric}
\eea
with
\bea
\Delta \equiv \sum_{\alpha=0}^4 X_\alpha \mu_\alpha^2\;,\quad
X_0 \equiv (X_1X_2X_3X_4)^{-2}\;,\quad
\mu_0^2\equiv 1- \sum_{a} \mu_a^2\;.
\eea
Dilaton and two-form field strength in ten dimensions are given by
\bea
\phi &=& \frac{1}{3} \log{\left(\rho^{-7/4} \Delta^{-9/8}\right)}
\;,\label{ansatz:dilF}\\[1ex]
F &=&  \Big( 2 \,
\rho^{5/9}\, g \,
\sum_{\alpha=0}^4 \left( X_\alpha^2 \mu_\alpha^2 - \Delta X_\alpha \right)
+   \rho^{5/9} \, g \,\Delta \,X_0\, \Big) \, \varepsilon_2
\nonumber\\
&&{}
+\frac{\rho^{13/9}}{2  g^{2}} \,
\sum_{a} X_a^{-2} d(\mu_a^2) \wedge \left(d\sigma_a +  g A^a\right)
\left(*_{2} \, F^{a} \right)
+\frac{\rho}{2 g} \,  \, \sum_{\alpha=0}^{4} X_{\alpha}^{-1} *_{2} dX_{\alpha} \wedge d(\mu_{\alpha}^{2})
\;,
\nonumber
\eea
with the two-dimensional volume form $\varepsilon_2$ and a two-dimensional Hodge star $*_{2}$
defined with respect to the two-dimensional metric $g_{\mu\nu}$\,.
All sums over $\alpha$ run from 0 to 4, the sums over $a$ run from 1 to 4.
Indeed, we have verified explicitly that with this reduction ansatz, the field equations of IIA supergravity
(\ref{L2A}) reduce to the field equations for the two-dimensional fields $\{g_{\mu\nu}, X_a, \rho, A_\mu^a\}$,
derived from the Lagrangian~(\ref{Ldilaton}) found in the previous section.
A convenient way to derive the proper form of the ansatz is by first embedding the solutions 
(\ref{const_rho1}), (\ref{const_rho2}) with constant
scalars and dilaton $\rho$ found above.
This fixes the ansatz up to the last term in the two-form field strength in (\ref{ansatz:dilF})
and the scaling symmetries of the ten-dimensional Lagrangian (\ref{L2A})
\bea
g_{MN} \rightarrow \lambda^{2} \, g_{MN}\;,\quad
F_{MN} \rightarrow \lambda \mu \, F_{MN}\;,\quad
\phi \rightarrow \phi - \frac{2}{3} \log \mu
\;,
\eea
with constant $\lambda, \mu$. Upon proper identification of the scaling
parameters $\lambda$, $\mu$ as functions of the dilaton $\rho$, one arrives at
the reduction ansatz (\ref{ansatz:metric}), (\ref{ansatz:dilF}), which can then be confirmed
by explicit calculation.

With (\ref{ansatz:metric})--(\ref{ansatz:dilF}) we have obtained the full
ten-dimensional embedding of the dilaton sector of the two-dimensional model~(\ref{LSO9}).
This is an indispensable tool for further holographic computations around the domain-wall
background (\ref{dow}) and other backgrounds in this truncation. Such applications to holography
of matrix quantum mechanics will be addressed in \cite{Ortiz:2013XX}.

\section{Rotating branes and Domain-Wall Black Holes}

\paragraph{Rotating D0 brane}

As a further application, let us consider the rotating brane solutions,
constructed in the appendix of~\cite{Cvetic:1999xp}, see also \cite{Harmark:1999xt}.
The case of relevance here, are the rotating $0$-brane solutions
of (\ref{L2A}).
In the limit of large brane charges, the ten-dimensional solution takes a form that falls into the
parametrization (\ref{ansatz:metric})--(\ref{ansatz:dilF}), explicitly given by
\bea
ds_2^{2} &=& (g r)^{7} h(r)^{-7/9} f(r) \, dt^{2} - h(r)^{2/9} f(r)^{-1} \, dr^{2} \;,\nonumber\\[.5ex]
A^{a}(r)&=&\frac{1-H_{a}(r)}{l_{a}} \,  \sqrt{2mg^5} \; dt
\;,\nonumber\\[.5ex]
\rho(r) &=& (g r)^{9/2} \,h(r)^{-1/2} \;,\nonumber\\[.5ex]
X_{a}(r) &=& h(r)^{-2/9} \,H_{a}(r) \;,
\label{rotating0}
\eea
with free constants $g, m, l_a$\,, and the functions
\bea
h(r)\equiv \prod_{a} H_{a}(r) \;,\qquad
H_{a}(r)\equiv \left(1+\frac{l_{a}^{2}}{r^{2}}\right)^{-1}\;,\qquad
f(r)\equiv 1-\frac{2 m \, h(r)}{ r^{7}} \;.
\eea
One may explicitly verify that the ansatz \eqref{rotating0} describes a solution
of the field equations of the two-dimensional theory derived in section~\ref{sec:truncation}.
In the massless limit $m\rightarrow0$, it preserves half of the supersymmetries.
W.r.t.\ the two-dimensional model, the metric $ds_2^2$ in (\ref{rotating0}) describes a domain-wall
black hole whose asymptotic curvature approaches (\ref{dow})
\bea
R = \frac{35}{2 \, r^2}
+ {\cal{O}} \left( r^{-23/9} \right)
\;,
\qquad
\mbox{for}\;\;r \rightarrow \infty
\;,
\eea
whereas at $r=0$ it behaves like
\bea
R = -\frac{7}{6} \,r^{-34/9}  \prod_{a} (l_a)^{4/9}
+{\cal{O}} \left( r^{-26/9} \right)
\;.
\eea
We may consider the near-horizon behavior of the two-dimensional metric
around $r=r_0$, where $r_0$ is the highest root
of the equation $f(r)=0$. To this end, we expand the coordinates according to
\begin{equation}
r \rightarrow r_0 + \epsilon \, r \;,\qquad
 t \rightarrow \rho_{0}^{-7/9} h_{0}^{-1/9} \, \frac{t}{\epsilon} \;,
\end{equation}
where
\begin{equation}
\rho_{0} \equiv (g r_0)^{9/2} h_{0}^{-1/2}\;,\quad
h_{0}\equiv \prod_{a=1}^{4} H_{a0} \;,\quad
H_{a0}\equiv \Big(1+\frac{l_{a}^{2}}{r_{0}^{2}}\Big)^{-1}\;.
\end{equation}
In the limit $\epsilon \rightarrow 0$, we then obtain the near-horizon AdS$_2$
configuration
\begin{equation}
ds^{2} = f_{0} \, dt^{2} - \frac{1}{f_{0}} \, dr^{2}\;,\qquad
F_{tr}^{a} = 2 g \, \rho_0^{-4/9} \left( \frac{{H}_{a0}^{2} \,
 \sqrt{ {H}_{a0}^{-1}-1}}{h_{0}^{4/9}}  \right)
 \;,
\end{equation}
with \bea f_{0}\equiv g^{2} \frac{\left(1+8 \sum_{a<b} {H}_{a0}
{H}_{b0}\right)}{2 \, \rho_{0}^{4/9} \, {h}_{0}^{4/9}} \, r^{2} \;,
\eea provided the constants $H_{a0}$ satisfy the following further condition \bea
\sum_{a=1}^{4} {H}_{a0}~=~1/2\label{HS} \;. \eea
This is the $C=0$ case of the solution \eqref{const_rho2} found
above. According to the embedding
(\ref{ansatz:metric})--(\ref{ansatz:dilF}), this solution
corresponds to a ten-dimensional warped product geometry AdS$_2
\times {\cal M}_8$\,.

\paragraph{Thermodynamics of non-rotating D0-branes}

Let us consider first the case when the angular momentum vanishes $l_{a}=0$.
The solution then has $A_{a}=0,$ $H_{a}=1$, $h(r)=1$, $X_{a}=1$, $\Delta =1$
and simplifies to%
\begin{equation}
ds_{10}^{2}=\rho ^{-7/36}\left[ \left( gr\right)
^{7}f(r)dt^{2}-f(r)^{-1}dr^{2}\right] -g^{-2}\,\rho ^{1/4}d\Omega _{8}
\;,
\end{equation}%
\begin{equation}
\exp (\phi )={\rho ^{-7/12},}\qquad F=-7\rho ^{5/9}\,g\varepsilon
_{2},\qquad \rho =(gr)^{\frac{9}{2}},\qquad f(r)=1-\frac{2m}{r^{7}}
\;,
\end{equation}%
where, again, the two-dimensional volume form $\varepsilon _{2}$ is defined
as: $\varepsilon _{2}=e\,dt\wedge dr$. The corresponding string frame metric%
\begin{equation}
\alpha ^{\prime }e^{\phi }ds_{10}^{2}=\alpha ^{\prime }\left( {\rho ^{-7/9}}%
\left[ \left( gr\right) ^{7}f(r)dt^{2}-f(r)^{-1}dr^{2}\right] -g^{-2}\,\rho
^{-1/3}d\Omega _{8}\right)
\;,
\end{equation}%
coincides with eq. (2.3) of \cite{Wiseman:2013cda} \footnote{%
It exactly coincides after the relabeling $r=U$ and $g^{-7}=a_{0}\lambda $
where $a_{0}=2^{7}\pi ^{\frac{9}{2}}\Gamma (\frac{7}{2})$.}.

The original asymptotically flat $p$-branes were constructed in \cite%
{Horowitz:1991cd} and their thermodynamics, as D$p$-branes, in the context of
the AdS/CFT correspondence was analyzed in \cite{Itzhaki:1998dd}. The
entropy of this configuration is given by the area law%
\begin{equation}
S=\frac{g^{-\frac{7}{2}}(2m)^{\frac{9}{14}}}{4G}{\rm Vol}(S^{8})
\;.
\label{Entropy}
\end{equation}%
Its temperature it is given by the inverse of the Euclidean period,%
\begin{equation}
\left( 2\pi T\right) ^{2}=-\left. \frac{g^{\mu \nu }\left( \partial _{\mu
}N\right) \left( \partial _{\nu }N\right) }{4N}\right\vert _{r^{7}=2m}=\frac{%
49g^{7}\left( 2m\right) ^{\frac{5}{7}}}{4}\;.  \label{Temperature}
\end{equation}%
where $N=g_{\mu \nu }K^{\mu }K^{\nu }$ and $K=\partial _{t}$ is the relevant
timelike Killing vector. The first law of black hole thermodynamics is%
\begin{equation}
\delta M=T\delta S=\frac{9{\rm Vol}(S^{8})}{16\pi G}\delta m
\;.
\end{equation}%
The energy of the spacetime is then given by the integration of the first law%
\begin{equation}
M=\frac{9{\rm Vol}(S^{8})}{16\pi G}m \;. \label{Energy}
\end{equation}%
Using (\ref{Temperature}) and (\ref{Energy}) it is possible to see that
\begin{equation}
M\sim T^{\frac{14}{5}}  \;,\label{MT}
\end{equation}%
in accordance with the results in the literature \cite{Wiseman:2013cda},
\cite{Itzhaki:1998dd}.

\paragraph{Thermodynamics of rotating D0-branes}

Let us now proceed to the analysis of the rotating solution
(\ref{rotating0}).
The metric
(\ref{ansatz:metric}) is in Boyer-Lindquist form and the eight
sphere is round when $r=\infty $. Therefore is possible
to identify the angular velocities of the horizon\footnote{%
A very good discussion about thermodynamical aspects of the Kerr black hole is \cite{Caldarelli:1999xj}.}%
\begin{equation}
\Omega _{i}=-g\,A_{t}^{a}(r_{+})\;,
\end{equation}%
therefore, the Killing vector that vanishes at the horizon is%
\begin{equation}
K=\partial _{t}-g\,A^{a}(r_{+})\partial _{\phi _{a}}\;,
\end{equation}%
where $2m=r_{+}^{7}h(r_{+})$ is the location of the horizon. Note that%
\begin{equation}
2\delta m=r_{+}^{7}h(r_{+})\left( \frac{7}{r_{+}}+\frac{h^{\prime }(r_{+})}{%
h(r_{+})}\right) \delta r_{+}+\frac{r_{+}^{5}h(r_{+})}{H_{a}(r_{+})}\delta
l_{a}^{2}\;.
\end{equation}%
The temperature is%
\begin{eqnarray}
\left( 2\pi T\right) ^{2} &=&\frac{g^{7}r_{+}^{7}h(r_{+})}{4}\left( \frac{14m%
}{r_{+}^{8}h(r_{+})}+\frac{2mh^{\prime }(r_{+})}{r_{+}^{7}h^{2}(r_{+})}%
\right) ^{2}  \notag \\
&=&\frac{g^{7}r_{+}^{7}h(r_{+})}{4}\left( \frac{7}{r_{+}}+\frac{h^{\prime
}(r_{+})}{h(r_{+})}\right) ^{2}
\;.
\end{eqnarray}%
The entropy is straightforward to compute%
\begin{eqnarray}
S &=&\frac{g^{-8}\rho
(r_{+})}{4G}{\rm Vol}(S^{8})=\frac{g^{-7/2}r_{+}\sqrt{2m}}{4G}{\rm Vol}(S^{8})
\;.
\label{entropy1}
\end{eqnarray}%
It follows that%
\begin{equation}
T\delta S=\frac{9}{16\pi G }{\rm Vol}(S^{8})\delta
m+\frac{{\rm Vol}(S^{8})}{8\pi G}\Omega _{a}\delta \left(
g^{-7/2}l_{a}\sqrt{2m}\right)
\;,
\end{equation}%
which allows to identify the energy $M$ and angular momenta $J_{a}$ of the
spacetime
\begin{equation}
M=9m\frac{{\rm Vol}(S^{8})}{16\pi G}\qquad J_{a}=-g^{-7/2}l_{a}\sqrt{2m}\frac{%
{\rm Vol}(S^{8})}{8\pi G}
\;.
\end{equation}%
It is possible to appreciate that the simple relation between the
energy density in the CFT and the temperature (\ref{MT}) does no
longer hold. Our expressions exactly coincide with the
thermodynamics in the near horizon limit of D0-branes given by eqs.\
(3.4a) and (3.4b) of \cite{Harmark:1999xt}\footnote{ It is necessary
to make the identification $h=g^{-1}$, $r^{7}_{0}=2m$ and
$V_{0}=1$.}.

\section{Entropy function}

We would like to understand the thermodynamics of the rotating D0
branes, in the extremal limit, from a two-dimensional point of view.
To this end we shall use the entropy function formalism introduced by Sen in~\cite{Sen:2005wa}
which also applies to the asymptotically AdS black holes in higher-dimensional
gauged supergravity~\cite{Morales:2006gm}.
The idea is
to take the scalar and gauge fields at some fixed value and the
metric to be AdS$_{2}$
\begin{equation}
ds^{2}=v^{2}\left( r^{2}dt^{2}-\frac{dr^{2}}{r^{2}}\right)\;, \qquad
F_{tr}^{a}=Q^{a}\;,
\end{equation}
which, when evaluated in (\ref{Ldilaton}), yield the following
Lagrangian
\begin{equation}
\mathcal{L}=-\frac{\rho }{2}+\frac{v^{-2}}{8}\rho
^{13/9}Q^{a}Q^{a}X_{a}^{-2}-\frac{v^{2}}{8}g^{2}\rho ^{5/9}V(X)\;.
\end{equation}
The entropy of the configuration is given by the extremal value of
the Legendre transform of the Lagrangian respect to $Q$
\begin{equation}
\mathcal{S}=2\pi\kappa(q^{a}Q^{a}-\mathcal{L})
\;,
\end{equation}
where the $2\pi$ factor is universal, the $\mathcal{L}$ is the two
dimensional Lagrangian (\ref{Ldilaton}), $q^{a}$ are the physical
charges and $\kappa$ is an overall multiplication constant defined
by the relation between the ten-dimensional action and the two
dimensional action. Indeed, the ten-dimensional Lagrangian,
(\ref{L2A}), has the canonical normalization in front of the Ricci
scalar $(16\pi G)^{-1}$, which allowed us to define the entropy in
the standard form (\ref{Entropy})
\begin{equation}
S=\frac{\mathcal{A}}{4G}
\;,
\end{equation}
where $\mathcal{A}$ is the black hole area.\footnote{For the sake of
simplicity and consistency with the original work
\cite{Ortiz:2012ib}, the two-dimensional Lagrangian has been written
without due normalization.} It follows that the constant $\kappa$ is
determined by the integral over the eight sphere of the ten
dimensional Ricci scalar
\begin{equation}
\kappa \rho (Re)_{2D}=\int_{S^{8}}(Re)_{10D}
\;.
\end{equation}
In this way we get
\begin{equation}
\kappa=\frac{{\rm Vol}(S^{8})g^{-8}}{4\pi G}
\;.
\end{equation}

The extremal value of the entropy function functional is given by
the equations 
\begin{eqnarray}
\kappa^{-1}\frac{\partial \mathcal{S}}{\partial Q^{a}}
&=&q^{a}-\frac{v^{-2}}{4}\rho
^{13/9}Q^{a}X_{a}^{-2}  \;,\label{AM 1} \\
\kappa^{-1}\frac{\partial \mathcal{S}}{\partial \rho }
&=&\frac{1}{2}-\frac{13v^{-2}}{72}\rho
^{4/9}Q^{a}Q^{a}X_{a}^{-2}+\frac{5v^{2}}{72}g^{2}\rho ^{-4/9}V(X)
\;,\label{AM 2} \\
\kappa^{-1}\frac{\partial \mathcal{S}}{\partial v} &=&\frac{v^{-3}}{4}\rho
^{13/9}Q^{a}Q^{a}X_{a}^{-2}+\frac{v}{4}g^{2}\rho ^{5/9}V(X)  \;,\label{AM 3} \\
\kappa^{-1}\frac{\partial \mathcal{S}}{\partial X^{b}} &=&\frac{v^{-2}}{4}\rho
^{13/9}Q^{b}Q^{b}X_{b}^{-3}+\frac{v^{2}}{8}g^{2}\rho
^{5/9}\frac{\partial V(X)}{\partial X_{b}} \;.  \label{AM 4}
\end{eqnarray}%
Equation (\ref{AM 1}) defines the electric charge $q^a$.
Equations (\ref{AM 2}) and (\ref{AM 3}) are not independent. Indeed,
is easy to see that
\begin{equation}
v^{-4}\rho ^{8/9}Q^{a}Q^{a}X_{a}^{-2}=2v^{-2}\rho
^{4/9}=-g^{2}V(X)~\Longrightarrow~ 2v^{2}\rho
^{-4/9}=Q^{a}Q^{a}X_{a}^{-2}\;.
\end{equation}%
The field equations for the scalar fields (\ref{AM 4})
can be interpreted as defining the value of the scalar fields
as function of the charges $q^a$ which is the
version of the attractor mechanism relevant for our setting.
From these equations, we get
\begin{eqnarray}
0 &=&-V(X)+\frac{1}{2}X_{b}\frac{\partial V(X)}{\partial X_{b}}
\;,
\end{eqnarray}
which is the condition (\ref{const_rho1}) for the $H_a$ adding up to $1/2$.
Together it is easy to get that the entropy of the two
dimensional, near extremal D0 branes is
\begin{equation}
\mathcal{S}=2\pi \kappa \frac{\rho }{2}=\frac{g^{-8}\rho}{4
G}{\rm Vol}(S^{8})
\;,
\end{equation}
in perfect agreement with our ten-dimensional result
(\ref{entropy1}). This result is very interesting, as it shows that
the entropy function formalism works very well even for objects that
have no area. Indeed, as the two-dimensional configurations
considered here have a zero-dimensional horizon, the value of their
entropies is far from clear. The fact that one can embed the
geometries in ten dimensions give us a prescription to compute its
thermodynamics, which turns out to coincide, in the near extremal
limit, with the entropy function formalism. This is a good
indication that the entropy function formalism can be safely used in
two dimensions, and could be extended even to cases where no
ten-dimensional embedding is known.

\section{AdS$_2 \times {\cal{M}}_8$ solutions with
non-vanishing axions}

Above, we have found
within the pure dilaton-sector of the two-dimensional model
the most general solution
with constant dilatons, thus lifting to ten-dimensional product geometries
AdS$_2 \times {\cal{M}}_8$. This results in the three-parameter family
of solutions (\ref{const_rho1})--(\ref{const_rho2}),
with the metric on ${\cal{M}}_8$ given by (\ref{ansatz:metric}).
Here, we sketch how the full $U(1)^4$ truncation (\ref{Ltrunc}) allows for
many more solutions of this type (with non-vanishing axions $\phi^a$),
although the structure of the equations
and their explicit solutions become much more complicated.

Let us start from the Lagrangian (\ref{Ltrunc})
and consider the equations of motion assuming all scalar fields to be constant.
The equations of motion for the auxiliary fields
$y^a$ determine the field strengths  ${F_{\mu \nu}}^{a} $.
In turn, the vector fields equations state that the $y^a$ are constants.
The field equation for $\rho$ fixes the AdS$_2$ radius.
Finally the remaining equations are
\begin{equation}
V_{\text{pot}} =0 \;,\qquad
\frac{\partial V_{\text{pot}} }{\partial H_a } = 0 \;,\qquad
 \frac{\partial V_{\text{pot}} }{\partial \eta_a } = 0
 \;,
 \label{eom_const}
\end{equation}
for the scalar potential $V_{\text{pot}}$ from (\ref{Vcomp}).
In total, these are nine algebraic equations for the 12 parameters
$H_a$, $y_a$, $\eta_a$ to be determined.
Moreover, we can explicitly solve the four last equations of (\ref{eom_const})
for the $H_a$, leaving us with five equations
for eight unknowns.
The counting suggests that there are various families of solutions to these equations
and even though the explicit equations are too complicated to allow for the general
explicit solution (collection of algebraic equations that are up to eighth order polynomials),
first numerical results support this idea.
Just as an example, we mention that an explicit solution can be found
upon further truncation the system
\begin{equation}
H_1\equiv H_3 \;,\; H_2\equiv H_4 \;,\qquad y_1\equiv y_3 \;,\; y_2\equiv y_4 \;,\qquad
\eta_1\equiv \eta_3 \;,\; \eta_2\equiv \eta_4\;.
\end{equation}
In this truncation, the equations can further be reduced to quadratic equations
and allow for the explicit solution
\bea
H_1 &=& \frac{1}{128} (43 - 5 \, \sqrt{33} ) \;,\quad
H_2 ~=~  \frac{1}{64} (25 + 9 \, \sqrt{33} ) \;,\nonumber\\
(y_1)^2 &=& 12 \, \big( 6 + \sqrt{33} \big)  \;,\;
(y_2)^2 ~=~ 2 \, \big( -1 +\sqrt{33} \big)  \;,\nonumber\\
(\eta_1)^2 &=& \frac{1}{8} (9 + \sqrt{33})\;,\; (\eta_2)^2 ~=~  \frac{1}{16}(1 + \sqrt{33})
\;.
\eea
with Ricci scalar given by
\bea
R = \frac{2^{2/3} \,3 \,\big(3815 + 759 \sqrt{33} \big) }{\Big(-205+131 \sqrt{33}\Big)^{8/9}}
\, \frac{g^2}{\rho^{4/9}} \simeq 143.27 \, \frac{g^2}{\rho^{4/9}}
\;.
\eea

It would certainly be interesting to get better control over the solutions of this type
in the full model (\ref{Ltrunc}) as well as on their higher-dimensional origin.
The latter may represent a major step towards finding the Kaluza-Klein ansatz for
the model with non-vanishing axion fields into the ten-dimensional theory.

\section{Acknowledgments.}

A.A. would like to thank Toby Wiseman and Dumitru Astefanesei for
insightful discussions. Research of A.A. is supported in part by the
Fondecyt Grant 11121187 and by the CNRS project ``Solutions exactes
en pr\'{e}sence de champ scalaire''.


\providecommand{\href}[2]{#2}\begingroup\raggedright\endgroup

\end{document}